# Off-stoichiometric variable doping for exceptional power factors in L2$_1$ $Fe_2VAlM_x$ (M=Ti, W) epitaxial thin films. †

Jose María Domínguez-Vázquez[a], Miguel Angel Tenaguillo[a], Ketan Lohani[a], Jose J. Plata[b], Antonio M. Marquez[b], Olga Caballero-Calero[a], Alfonso Cebollada[a], Andrés Conca[a*], Ernst Bauer[c], and Marisol Martín-González[a]

We show that the addition of Ti or W to stoichiometric L2$_1$ $Fe_2VAl$ thin films by sputter codeposition produces off stoichiometric thin film alloys with superior thermoelectric properties than their stoichiometric counterparts. Ti incorporation induces p-type semiconducting behavior, while W incorporation shifts the material toward n-type, hereby enabling simultaneous tuning of both carrier types within a single parent ($Fe_2VAl$) material system, making it highly desirable for thermoelectric devices. The introduction of both Ti and W partly substitutes V in the stoichiometric compound. The partial substitution of V in the stoichiometric alloy allows fine-tuning the band structure of the system and transport properties. With this approach we obtain exceptional maximum power factor values for p and n-type films of 1300 μW/m·K² and 2100 μW/m·K² ,respectively, yielding maximum figures of merit *zT* of 0.07 and 0.14, respectively.

[a] *Instituto de Micro y Nanotecnología, IMN-CNM, CSIC (CEI UAM+CSIC), Isaac Newton 8, E-28760 Tres Cantos, Madrid, Spain*
[b] *Departamento de Química Física, Facultad de Química, Universidad de Sevilla, Seville, Spain*
[c] *Institute of Solid-State Physics, Technische Universität Wien,Wiedner, Hauptstraße 8-10, Vienna, 1040, Austria*
**Corresponding author: andres.conca@csic.es*
† Supplementary Information available

## Introduction

Thermoelectric materials have the unique ability of transforming thermal gradients into voltage differences. Given the increasing challenges of energy generation, they are well-placed as candidates for increasing global energy efficiency by recovering unused or wasted thermal energy, offering advantages over other devices such as a high reliability and lack of moving parts. They also offer a solution for powering the growing number of Internet of Things (IoT) devices. Thermoelectric efficiency is evaluated through the dimensionless figure of merit (zT), which is defined as $zT=S^2\cdot\sigma\cdot T/\kappa$ where S, σ and κ are the Seebeck coefficient, electrical conductivity and thermal conductivity, respectively, and T is the temperature. Therefore, high-performing thermoelectric materials should exhibit high Seebeck coefficient and electrical conductivity, often combined in the form of the power factor $PF=S^2\cdot\sigma$, and low thermal conductivity.

State-of-the-art thermoelectric materials currently include bismuth telluride ($Bi_2Te_3$)[1–3], lead telluride (PbTe)[4], silicon-germanium (SiGe) [5] or metal selenides ($Ag_2Se$ [6–8], $Cu_2Se$ [9] or SnSe [10,11] for instance). While these materials offer excellent thermoelectric performances over a wide range of temperatures, they all have some drawbacks that limit their mass application such as the scarcity, toxicity or high cost of their constituent elements. To address these issues, half and full Heusler materials have emerged as promising alternatives as some of them are composed of earth-abundant and non-toxic elements. Half and full Heusler alloys have formulas of XYZ and $X_2YZ$, respectively, with X and Y being transition metals and Z a main group element. Among all full Heusler alloys $Fe_2VAl$ stands out as one of the best for thermoelectric applications: its narrow-gap band structure, combined with doping versatility using elements such as W, Ta, Ti or Si results in high values of the power factor (PF) and the figure of merit (*zT)* near room temperature. Furthermore, such substitutions allow tuning of the material toward either p- or n-type conduction, facilitating complementary functionality within a single base material, a highly desirable aspect for thermoelectric device applications.

Ideally, full Heusler alloys are composed of four interpenetrating FCC lattices forming the so-called L2$_1$ structure, with each element occupying specific lattice sites. However, chemical disorder, also referred to as anti-site defects, is highly common, forming crystalline phases known as B2 or A2, in which different elements occupy indistinctively common lattice sites. In a previous work, we demonstrated the effect that L2$_1$ ordering has on the band structure of undoped $Fe_2VAl$ and, subsequently, on the thermoelectric properties[12]. Density functional calculations show that the chemically ordered L2$_1$ structure has a narrow-gap electronic structure whereas the A2 phase is predicted to be metallic. The exact nature of the gap (semiconductor vs. compensated semimetal) remains sensitive to the exchange-correlation functional employed, and our r2SCAN treatment, which substantially improves upon standard GGA, predicts a gap value of 0.18 eV which aligns with experimental optical measurements. Regarding thermoelectric performance, $L2_1$ $Fe_2VAl$ is found to have a twofold increase in the Seebeck coefficient with respect to the disordered counterparts.

Partial substitution of V by W in this chemically ordered system paves a way for further improvement, a threefold increase in the Seebeck coefficient and an almost tenfold increase in the thermoelectric power factor has been recently obtained in $L2_1$ $Fe_2V_{0.8}W_{0.2}Al$ thin films compared to those in the B2- or A2 phases[13]. Even more, the W-doped $L2_1$ alloy showed more than a fourfold increase on the maximum *zT* with respect to undoped $L2_1Fe_2VAl$, proving synergistic effects of W doping and chemical order.

In general, partial substitution has been a widely used approach to improve the thermoelectric performance through doping optimisation in of $Fe_2VAl$ alloys, but most of them on bulk specimens. For instance, Mikami *et al.[14]* and Hinterleitner *et al.*[15] performed W substitution optimization, both obtaining maximum values for *zT* around 0.2 in the 70-120 °C temperature range. Similarly, Masuda *et al.*[16] and Garmroudi *et al.*[17] used Ta and Ta-Si doping, respectively, obtaining maximum values for *zT* of 0.3 at 130 °C and 0.33 at 25 °C.

All these optimization routes can be exploited in a straightforward way if a thin film deposition approach is considered. For example, chemical order is optimized thanks to surface diffusion mechanisms as the film is grown, which allows the atoms to find the right lattice position at lower temperatures compared to bulk annealing. On the other hand, homogeneous and variably controlled doping is easily achieved by co-deposition from independent material sources. Last, but not least, the intrinsic reduced dimensionality nature of a thin film implies a convenient reduction of the thermal conductivity.

In this scenario, some alternatives have been explored. Most of these works share that the best Seebeck coefficients, power factor and *zT* values are found on films that show $L2_1$ ordering, with the values of the mentioned quantities scaling with the degree of chemical order. Notably, the work of Hinterleitner *et al.[18]* reported a *z*T of 6 from a sputtered $Fe_2V_{0.8}W_{0.2}Al$ film with posterior annealing. Using a similar experimental procedure, Yalev *et al.* [19] reported a zT value of ~3.9, however, this last work revealed that possible substrate artifacts could affect the observed thermoelectric properties and the measurement conditions continue to be discussed in the community. Other works, like Kurosaki *et al*.[20] or Machda *et al*.[21] have studied doped $Fe_2VAl$ thin films synthesised by sputtering, including chips of Al, W or V on the full Heusler targets to vary the film composition and doping content, obtaining maximum power factors of 1000μW/m·K² and 1600μW/m·K², respectively, and maximum figures of merit of 0.12 and 0.16, respectively. Works of Kudo *et al.*[22] and Bourgault *et al.*[23] used co-deposition techniques to synthesise off-stoichiometric undoped $Fe_2VAl$ combined with high deposition temperatures. They obtained maximum power factors of 1260 μW/m·K² and 5600 μW/m·K² respectively, which exceed notably the average power factors of $Fe_2VAl$.

Clearly, it is well established that off-stoichiometry, obtained with controlled composition variation, combined with a high degree of $L2_1$ order leads to improved thermoelectric properties. However systematic studies that combine doping optimization of $Fe_2VAl$ thin films with elements, like W or Ti, with high $L2_1$ order are scarce. Here we present a combined theoretical and experimental approach where variable off-stoichiometric doping mechanism, combined synergistically with the presence of $L2_1$ chemical order, is discussed and employed to optimise the thermoelectric properties of $Fe_2VAl$ epitaxial films. Our off-stoichiometric doping strategy combined with chemical order results in an increase of electrical conductivity and Seebeck coefficient with respect to undoped $L2_1$ $Fe_2VAl$ and stoichiometrically doped $Fe_2V_{0.8}W_{0.2}Al$ films, achieving a more than two-fold and four-fold increase in power factor for p-type and n-type doping respectively with respect to $L2_1$ $Fe_2VAl$. Additionally, a decrease in thermal conductivity is observed, especially notable in W-doping, resulting in a more than three-fold and six-fold increase in the figure of merit for p-type and n-type respectively with respect to undoped $L2_1$ $Fe_2VAl$ thin films.

## Experimental details

A series of 150-330 nm thin films of $Fe_2VAlM_x$ (M=Ti, W) was deposited at 900°C in a co-deposition configuration, the power applied to the doping element magnetron was varied to achieve different doping concentrations. Each deposition was simultaneously performed on two substrates, MgO (1 0 0) and $Al_2O_3$ (1 1 $\bar{2}$ 0) to promote different crystalline orientations. The deposition was carried out in a UHV chamber (base pressure ~$10^{-9}$ mbar) utilizing DC magnetron sputtering. A stoichiometric commercial $Fe_2VAl$ target (Mateck GmbH) was sputtered at 80W and $3\cdot10^{-3}$ mbar Ar pressure, yielding a deposition rate of 1.94 nm/min, the power applied to the Ti magnetron ranged between 9W and 40W while the one applied to the W magnetron was between 1W and 5W. The doping concentration x was measured via EDX using a FEI VEIROS 460. Deposition temperature was measured in situ using a calibrated thermocouple located in the sample holder. The structure was characterized by X-ray Diffraction (XRD) measurements performed on a Bruker D8 Discover four-circle diffractometer with a microfocus X-ray source (IµS) (Cu $K\alpha_1$) and an Eiger2 2D detector. For the measurements of electrical conductivity and carrier concentrations at varying temperature a lab-made Hall measurement system present at TU Wien was used, in this system the magnetic field is ramped up to 10T for each temperature measurement. For Seebeck measurements with varying temperature a commercial ULVAC ZEM-3 system present at TU Wien was used.

Single parabolic calculations were made with the SeeBand tool[24], fitting was made fixing the effective mass ratio of electrons and holes of $m_2/m_1$=0.89 and changing solely the Fermi level position and Energy band gap. For the Ti doped film, the best fit was found for an energy band gap of 0.01eV and a Fermi level displacement of 0.09eV into the valence band. For W doped film fitting, a band inversion of -0.03eV and a Fermi level displacement of 0.06eV into the conduction band were used.

The thermal conductivity was measured in the out-of-plane direction using the time-domain thermo-reflectance (TDTR) method at various temperatures, utilizing the Front/Front configuration. The measurements were performed with a PicoTR system (PicoTherm), employing a pump and probe laser with wavelengths of 1550 nm and 750 nm, respectively. Both lasers feature a pulse duration of 0.5 ps, and the laser pulses were applied to the film (after depositing a thin Pt layer of 100 nm on top of it) within a time interval of 50 ns. From these measurements, the thermal diffusivity is obtained, and from that and knowing the different parameters of the film and substrate density of 6.41 ± 0.1 g/cm$^3$ and 7.05± 0.1 g/cm$^3$ for Ti and W doped respectively was estimated through X-ray reflectometry and heat capacity of 523 J/kg·K and 456 J/kg·K respectively for Ti and W doped alloys, estimated via rule of mixtures using bulk material values [25]. To account for the sensitivity of the multilayer thermal model in the Cp estimations alongside inherent variances in transducer thickness and layer densities, a standard procedure of multiple measurements was applied under identical conditions, yielding a conservative cumulative uncertainty of ±15% for the reported thermal conductivity values. Thermal boundary resistance between the Pt transducer and the film was determined to be in the range of 1.2-4 m$^2$K/W and was accounted for in the TDTR fitting.

## Computational details

Ground states were fully relaxed using the VASP package[26] with projector-augmented wave (PAW) potentials[27]. To correct the band gap underestimation typical of GGA functionals, single-point energies and wave functions were subsequently computed using the meta-GGA r2SCAN functional proposed by Furness et al.[28] Core and valence electrons were selected according to the standards proposed by Calderon et al.[29]. A plane-wave kinetic energy cut-off of 500 eV was employed to minimize Pulay stress errors, along with a dense k-point mesh of 8000 k-points per reciprocal atom to accurately describe the potential energy surface and the ground state wavefunction.

Potential metallic V-W segregated phases were identified by computing the convex hull of the Fe-V-Al-W system using the OQMD database[30]. Simulated X-ray diffraction patterns were obtained using the VESTA code[31]. Defect formation energies were calculated using Fe2VAl and the most stable pure phases as references. Doped models with varying Ti and W concentrations were constructed using Special Quasirandom Structures (SQS [32]) as implemented in the ICET code[33]. SQS cells provide a good approximation of random alloys, as their cluster vectors closely resemble those of truly random alloys. To account for long-range disorder, a 2×2×2 supercell of $Fe_2VAl$, based on the conventional unit cell and containing 128 atoms in total, was constructed. Geometry and lattice vectors were fully relaxed until the forces on each atom were below $10^{-3}$ eV Å$^{-1}$. Wave-function convergence was achieved when the energy difference between successive electronic steps fell below $10^{-7}$ eV, incorporating an additional support grid for the evaluation of augmentation charges to minimize force-related noise. To compare the electronic structures of the $L_{21}$ phase and the doped structures, band structure unfolding[34] was performed using the easyunfold code[35].

## Results and discussion

In this work, high-temperature sputter co-deposition from independent stoichiometric $Fe_2VAl$ and Ti or W targets was employed to obtain off-stoichiometric $L2_1$ chemically ordered $Fe_2VAl$ alloys with variable Ti and W dopant concentrations, which resulted in p- and n-type conduction, respectively. Figure 1 sketches the deposition configuration. $Al_2O_3$(1 1 $\overline{\mathbf{2}}$ 0) and MgO (0 0 1) substrates maintained at 900ºC were indistinctively used, as they both lead to $L2_1$ $Fe_2VAl$ epitaxial films of identical properties as was shown previously [12,13]. Epitaxial growth was systematically verified by X-ray Diffraction (XRD) measurements. The presence of $L2_1$ chemical order was assessed through the observation of the (1 1 1) diffraction peak in off-specular XRD measurements [12,13] (off-specular scans are included in the supporting information). The dopant amount and therefore the degree of off-stoichiometry was simply achieved by leaving fixed the power of the stoichiometric $Fe_2VAl$ target and varying that of the dopant (a calibration plot of doping concentration measured via EDX with respect to the electrical power applied to dopant magnetron is shown in the supporting information).

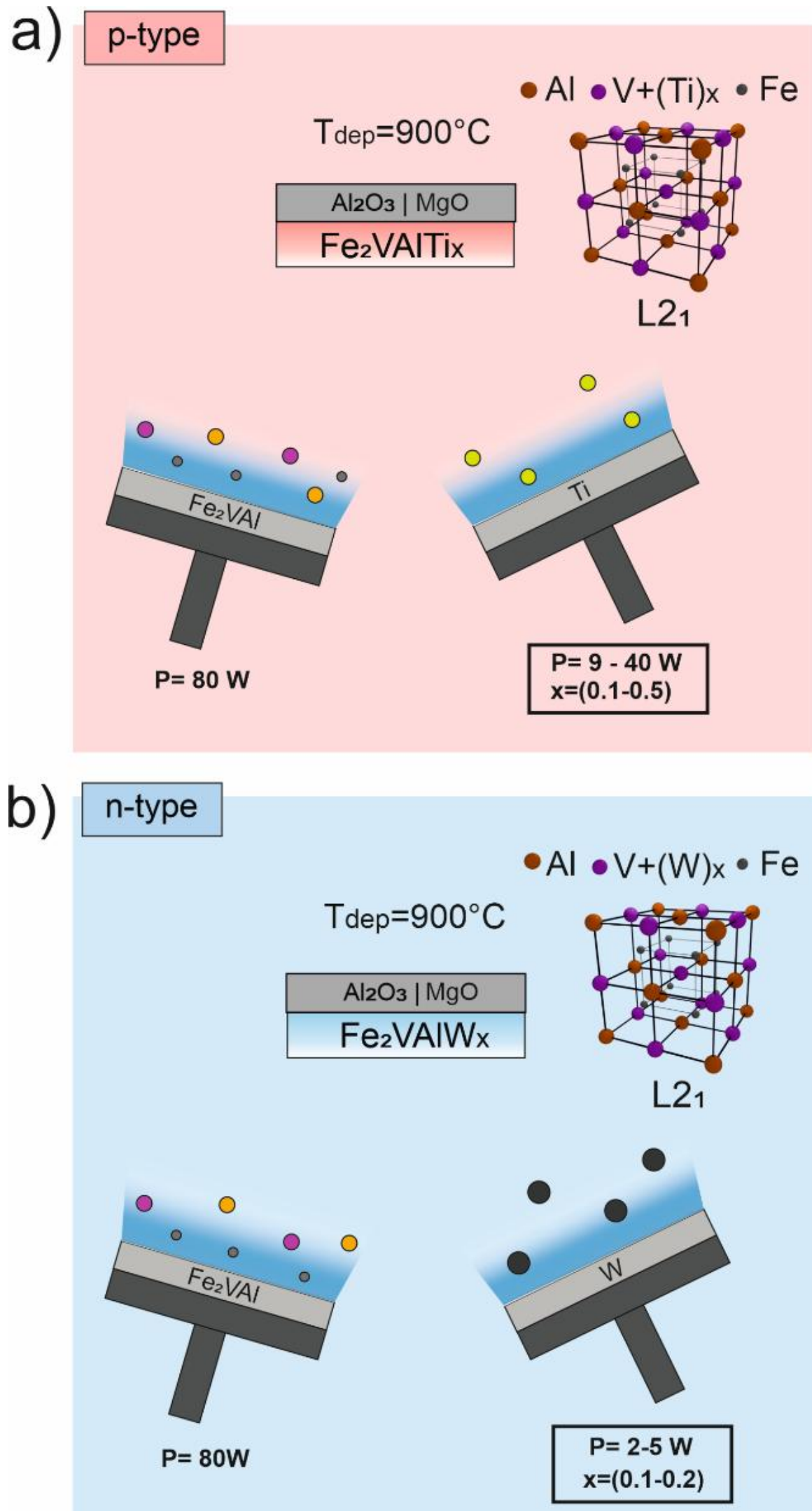

Figure 1: $Fe_2VAlM_x$ co-deposition configuration for depositing a) p-type (M=Ti) and b) n-type (M=W) $Fe_2VAl$-based $L2_1$ films on two crystalline orientations with ranging amounts of doping concentration.

Comprehensive X-ray characterization is summarized in Figure 2, showing for simplicity the results for films grown on $Al_2O_3(11\bar{2}0)$ substrates. $\theta$-$2\theta$ scans reveal the evolution of the system as higher amounts of dopant are introduced in $Fe_2VAlM_x$ (M=Ti,W). In the case of Ti-doped alloys, shown in Fig. 2a), single phase $L2_1$ Heusler crystallization was observed for the complete range of doping concentrations. At the specular condition ($\Psi$=0°), apart from the two diffraction peaks of the $Al_2O_3$ substrate, the only observed peaks are the (2 2 0) and (4 4 0) alloy peaks. Especially better observed in the (4 4 0) reflection, whose magnification is shown in figure 2 b), there is a continuous shift towards lower scattering angles as the amount of dopant is increased, indicative of an increase in the lattice parameter with increasing amount of Ti. We have explored formation energies for interstitial or Ti substitution of Fe, V or Al including the displacement of any of these three elements to the interstice, finding that most favourable scenario is the substitution of V by Ti (table with formation energies for all cases shown in the SI). This is consistent with the gradual evolution of the out of plane lattice parameter, from that of stoichiometric $Fe_2VAl$ for small amounts of Ti added, towards bulk $Fe_2TiAl$ [36], for increasing amounts of Ti. This linear and smooth behaviour, characteristic of a system following a Vegard's law dependence of the lattice parameter with dopant amount, confirms the gradual substitution of V by Ti in the lattice. Additionally, Special Quasirandom Structures (SQS) structures have been generated to model this Heusler lattice with different Ti concentration. The computed lattice parameters (at 0K) have been corrected to 20 °C using the thermal expansion coefficient of stoichiometric $Fe_2VAl$ [37]. This correction is applied uniformly across all compositions; at the highest Ti contents ($x \geq 0.38$), where the alloy composition approaches $Fe_2TiAl$, the thermal expansion coefficient may differ slightly from that of the parent compound, introducing a minor systematic uncertainty estimated to be below 0.8% in the corrected lattice parameter. Using this approach, lattice parameters obtained at 300 K with varying Ti contents are in good agreement with the experiments within an average underestimation of 0,05 Å, as it is shown in Figure 2 c). The absence of additional diffraction peaks indicates that the leftover V upon Ti substitution is likely located at the grain boundaries or forming extremely small grains with low volume fraction that do not manifest in the diffraction pattern.

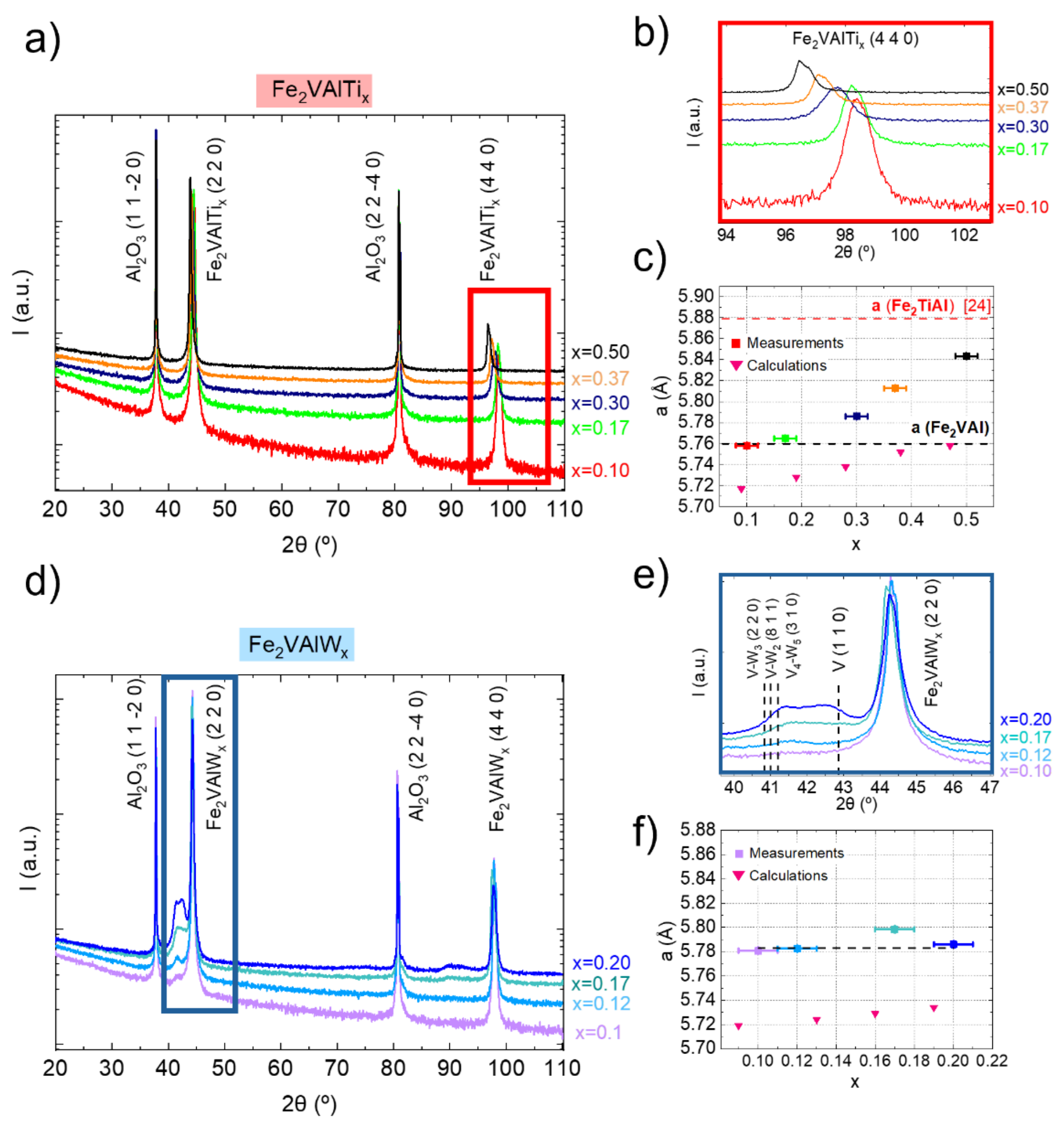

Figure 2: Structural X-ray characterisation. $\theta$- $2\theta$ scans of Ti-doped a) and W-doped d). b) and e) panels portray a zoom in a specific range of interest of a) and d) respectively. Lattice parameters of Ti-doped c) and W-doped f) films for the complete range of compositions. Error bars of lattice parameters are smaller than the marker size.

On the other hand, W addition produces a different effect on the system. As Figures 2 d) and e) show, irrespective of the amount of W added, the alloy (220) and (440) diffraction peaks are unaltered, while two additional Bragg-peaks gradually appear at lower scattering angles. Considering formation energies for interstitial or W substitution of Fe, V or Al with the displacement of any of these three elements to the interstice positions, we find that most favourable scenario is the substitution of V by W (table with formation energies for all cases shown in the supporting information). Also, SQS structures generated to model this Heusler alloy with various W concentration present lattice parameters (again at 0K and then corrected by $Fe_2VAl$ thermal expansion coefficient) are hardly changed and similar to pure $Fe_2VAl$ regardless the amount of W added. This lattice parameter evolution is in good agreement with the experimentally observed one, again with an underestimation of 0.06 Å, as shown in Figure 2 f). The two additional diffraction peaks whose intensity gradually increases with the amount of W correspond to V and VWx crystallites. The increasing intensity of the V crystallisation peak with increasing amount of W added to the alloy indicates that W atoms are effectively displacing V from the alloy. Thermodynamic stable compounds such as $V_2W_5$, $VW_2$, and $VW_3$ present diffraction peaks in this region, which are due to excess W and leftover V from the alloy. To experimentally discard major aggregations or inhomogeneities of Ti, W or V in the films, Energy-Dispersive X-ray Spectroscopy (EDX) maps were performed and a homogeneous nature was observed in both Ti-doped and W-doped films (EDX maps can be found in the supporting information).

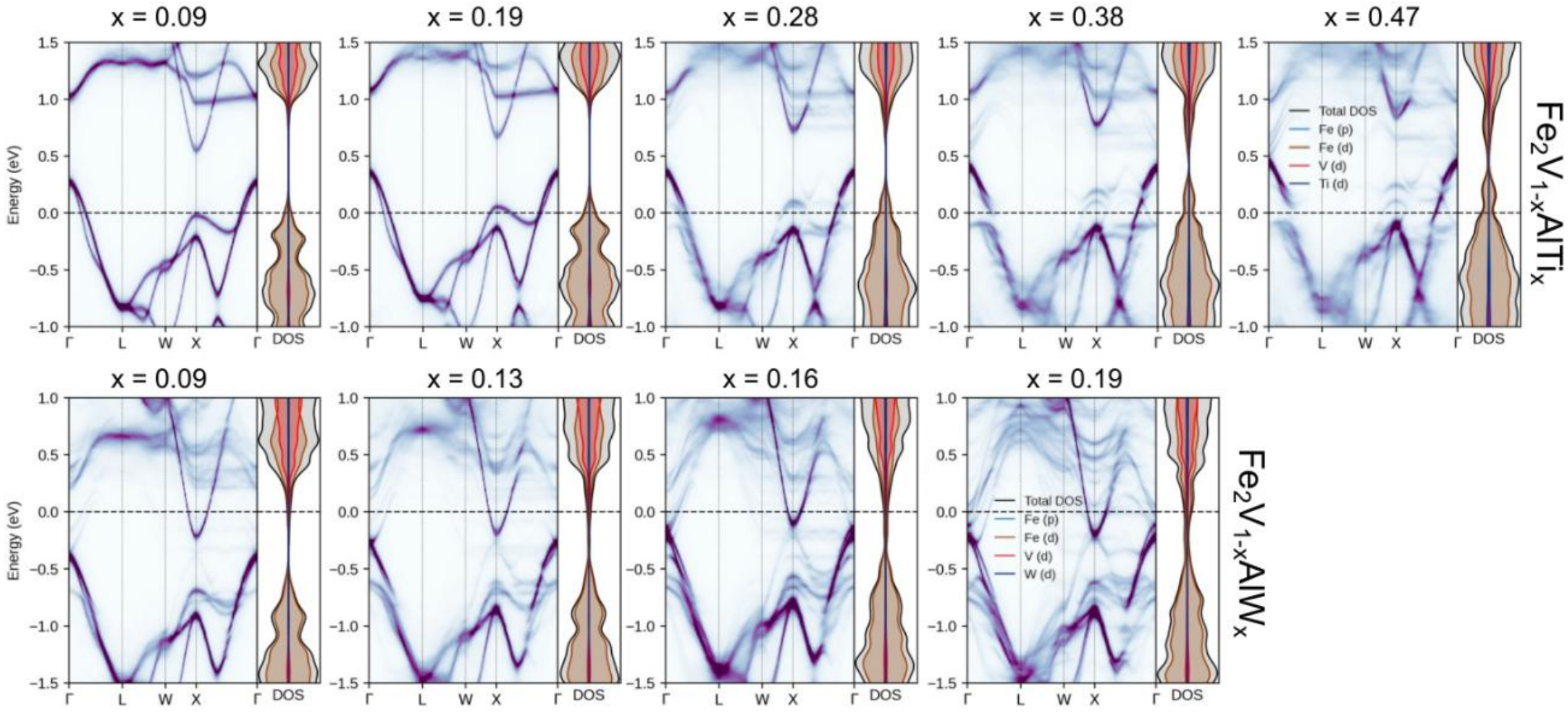

Figure 3: Unfolded band structures and density of states of stoichiometric Ti-doped and W-doped supercells. Due to their similarity and for simplicity, both minority- and majority-spin unfolded band structures are plotted in the same figure. The DOS plots also include atoms projections with the largest contributions to the total DOS around the Fermi level.

Theoretical band structure and Density of States (DOS) calculations, predict different band feature modifications for Ti and W doping. Figure 3 depicts unfolded band structures and DOS for stoichiometrically doped $Fe_2V_{1-x}M_xAl$ (M=Ti, W) alloys. Regarding Ti doping, when introduced in low amounts, a p-type rigid band shift is produced, increasing the density of states and the slope of the DOS at the Fermi level. When higher Ti contents are introduced, the slope of the density of states at the Fermi level is reduced without a major increase in the density of states caused by notable changes in the band structure in the X point. On the other hand, W doping, in concentrations of x=0.13 or higher, produces important alterations on the $Fe_2VAl$ band structure. Instead of a rigid band shift, a band inversion occurs at the Γ point, featuring complementary upshifts of valence band states and downshifts of conduction band states that yield a semi-metallic band structure. Electrical transport characterization is summarized in Figure 4. Here, Hall resistivity $R_{xy}$ measurements at 20 °C for representative doping concentrations of Ti and W are depicted along with obtained carrier concentrations for a temperature range from room temperature to 250 °C. For Ti-doped films, a linear behaviour of Hall resistivity with magnetic field is observed (Figure 4 a), indicating a predominant single carrier behaviour in electronic transport. Further, increasing Ti concentration results in a reduction of the Hall coefficient, proving an increment in carrier concentration. Figure 4 b) shows that this relation between Ti concentration and carrier concentration is maintained, generally, for the whole temperature range. Also, all films show a positive tendency of carrier concentration with temperature.

In a remarkable contrast with Ti-doped films, Hall resistivity of $Fe_2VAlW_x$, shown in Figure 4 c), shows a strong non-linearity with high magnetic fields, which is a typical footprint of multi-carrier electrical transport. In order to extract electrical transport information from Hall resistivity versus B data, a two-carrier model was used to analyse the results [38,39] (details of the analysis process can be found in the experimental details section):

$$\boldsymbol{R_{xy}} = \frac{\boldsymbol{n_1 q_1}\mu_1^2 + \boldsymbol{n_2 q_2}\mu_2^2 + (\boldsymbol{n_1 q_1} + \boldsymbol{n_2 q_2})\mu_1^2\mu_2^2\boldsymbol{B^2}}{(\boldsymbol{n_1 q_1}\mu_1 + \boldsymbol{n_2 q_2}\mu_2)^2 + (\boldsymbol{n_1 q_1} + \boldsymbol{n_2 q_2})^2\mu_1^2\mu_2^2\boldsymbol{B^2}}\boldsymbol{B} \qquad (1)$$

Where $R_{xy}$ is the hall resistivity, $n_1$ and $n_2$ are the carrier concentrations of majority and minority carriers respectively, $q_1$ and $q_2$ are the electric charges of majority and minority carriers respectively, $\mu_1$ and $\mu_2$ are the mobility of the majority and minority charge carrier respectively and B is the magnetic field.

Figure 4 d) shows that the majority of charge carriers are electrons, with concentrations of $10^{21}$-$10^{22}$ $cm^{-3}$ that increase with higher levels of tungsten doping. This experimental range is in fair agreement with theoretical predictions when the W content and optimized volume of the SQS supercell are considered (2-4 $10^{21}$ $cm^{-3}$ range). As our previous results on stoichiometrically doped $Fe_2V_{0.8}W_{0.2}Al$ demonstrate [13], this alloy exhibits asymmetric electrical transport behaviour, with flat band electrons acting as the majority carriers and holes as the minority carriers. The electron concentration is three orders of magnitude higher than hole concentration, whereas holes have much higher mobilities (carrier mobilities for Ti-doped and W-doped alloys are included in the supporting information). Notably, this double carrier behaviour is in good agreement with theoretical calculations presented in Figure 3, where valence band states are upshifted closer to the Fermi level at the Γ point.

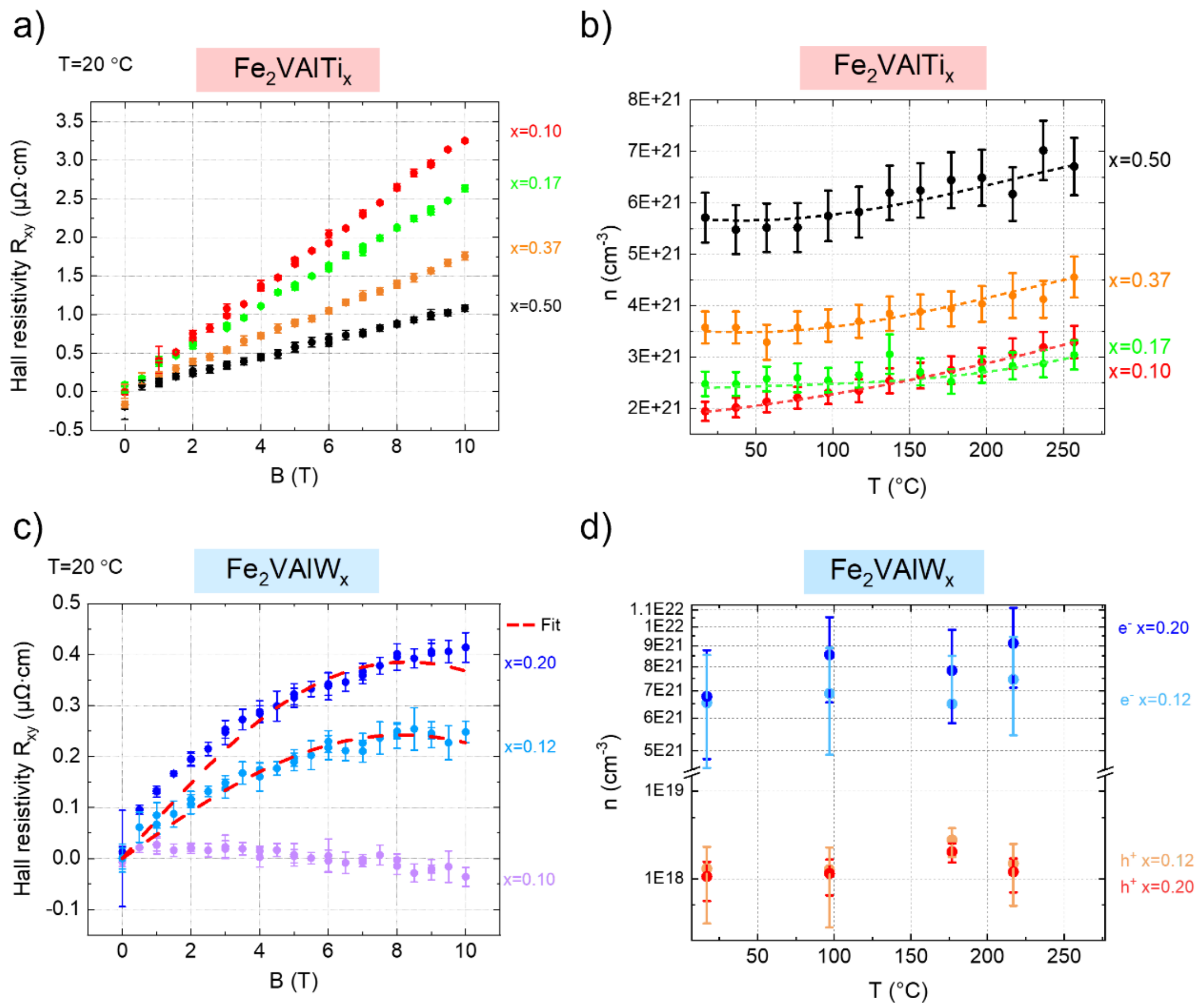

Figure 4: Electrical transport characterisation. Magnetic field-dependent Hall resistivity at T=20 °C of Ti-doped a) and W-doped c) and Temperature dependence of carrier concentration of representative films with ranging doping concentrations of Ti b) and W d).

Figure 5 shows temperature-dependent Seebeck coefficient for the entire range of Ti and W doping, along with measurements of undoped $Fe_2VAl$ and stoichiometrically doped $Fe_2V_{0.8}W_{0.2}Al$ obtained from our previous works [12,13]. The asymmetric electronic transport for the W doped alloys explains here the observed negative Seebeck coefficient in spite of positive Hall coefficients. In a two-carrier system, the Hall coefficient is weighted by $n_i\mu_i^2$ (quadratic in mobility), whereas the Seebeck coefficient is weighted by the individual conductivities $\sigma_i S_i/(\sigma_1+\sigma_2)$, where $\sigma_i=n_i e\mu_i$ scales linearly with mobility. Consequently, the high-mobility hole minority carriers dominate the sign of $R_H$ (positive), while the numerically dominant electron majority carriers, with lower mobility, determine the sign of $S$ (negative). This decoupling of the Hall and Seebeck signs is a direct fingerprint of the W-induced band inversion.

Maximum Seebeck coefficient values obtained on Ti- and W-doped samples surpass notably the maximum values obtained for undoped $Fe_2VAl$. For Ti-doped alloys a maximum Seebeck coefficient of 84 µV/K is obtained for a doping concentration of x=0.37 at a temperature of 180 °C. These results can be explained by analysing the changes in the unfolded band structure of Ti-doped supercells at different dopant contents. For Ti samples, the doping concentration does not strongly modify the band structure or the DOS at low Ti concentrations. However, there is a shift in the Fermi energy that opens the door to tuning the S coefficient, modifying dopant content. A larger DOS around the Fermi level, and consequently higher S coefficients, are obtained between x=0.28 and x=0.38, which agrees with the measurements. On the other hand, W-doped alloys show the maximum absolute Seebeck coefficient of 65 µV/K for the alloy with lowest amount of W, with a concentration of x=0.1, at a temperature of 270 °C. This smaller effect of W doping, with respect of Ti doping, on the Seebeck coefficient values can be explained by comparing their band structures. While Ti doping shifts the Fermi level, W doping does not significantly change the Fermi level but rather modifies the band structure. The greater the W doping concentration, the more W states appear at the bottom of the conduction band, mostly at the Γ point. Consequently, the DOS around the Fermi level does not change significantly, reducing the impact on increasing the Seebeck coefficient. Furthermore, a double parabolic band model[24] was used to fit Seebeck coefficient versus temperature data. The double band model fitted best the Ti-doped films data when a narrow gap and a Fermi shift into the valence band is introduced, and conversely, the best fit of the W-doped films data was found when a band inversion of 0.03eV and a Fermi level shift towards conduction band was introduced. This result is in good agreement with the theoretical band structures presented in Figure 3 while fitting well the Seebeck data, supposing an additional confirmation of the calculated rigid band shift for the Ti doping case and band inversion for the W doping case. Additionally, the crystallisation of V and VWx metallic aggregates observed by XRD can also reduce the enhancement of *S*. Nevertheless, the film with x=0.1 surpasses notably the maximum Seebeck coefficient obtained in undoped $Fe_2VAl$ and stoichiometrically doped $Fe_2V_{0.8}W_{0.2}Al$.

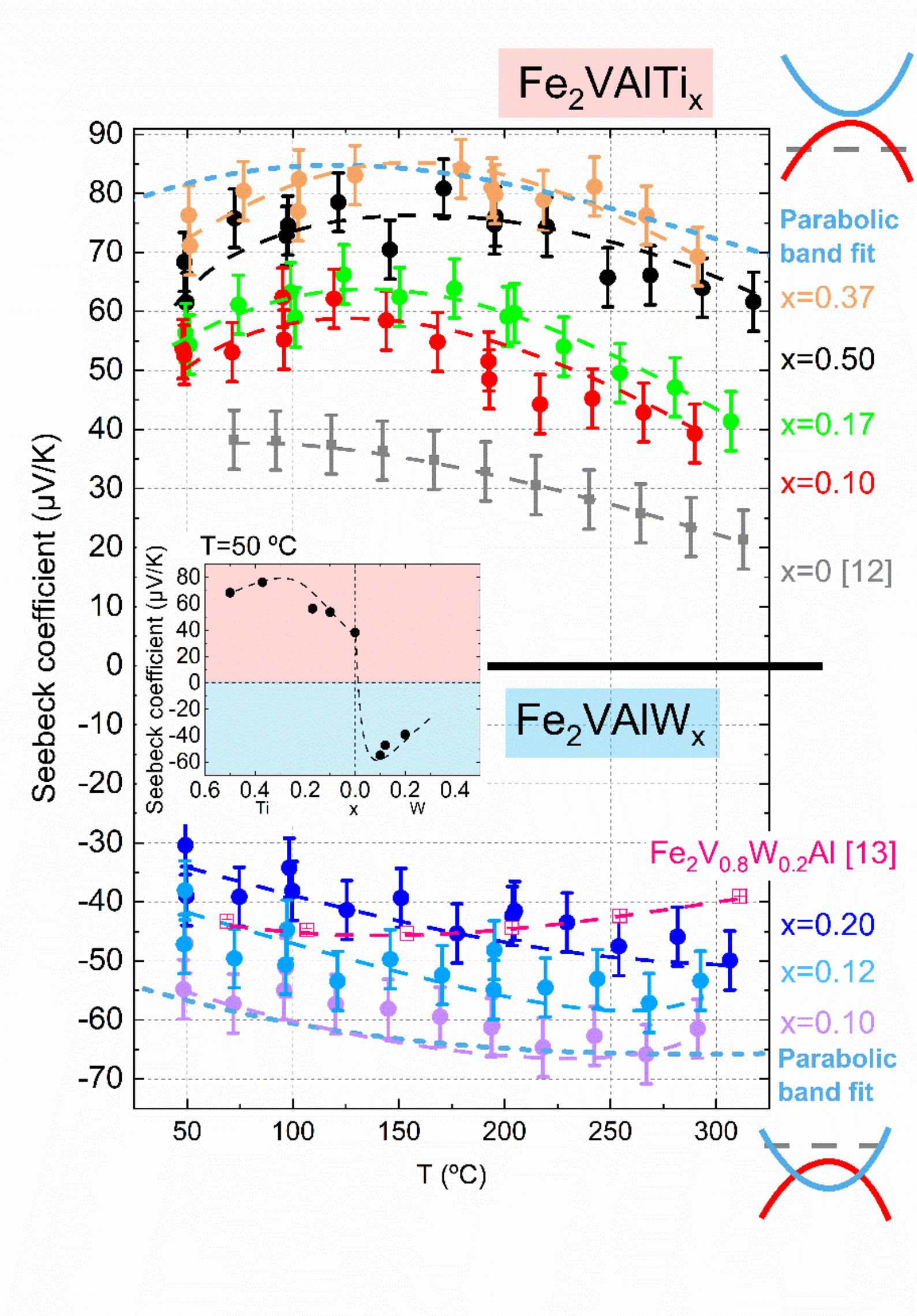

Figure 5: Temperature-dependent Seebeck coefficient of representative films with ranging amounts of Ti and W doping. Parabolic band fitting of data of $Fe_2VAlTi_{0.37}$ and $Fe_2VAlW_{0.10}$ is represented with the light blue dashed line along with a scheme of the band structure used for fitting. Inset: Seebeck coefficient at 50 °C versus Ti and W concentration.

In the inset to Figure 5 we show the compiled values of Seebeck coefficient at 50 ºC for Ti and W doping. The obtained curve qualitatively resembles the typical Seebeck coefficient vs valence electron concentration dependence [40,41], however, Ti doping exhibits a maximisation at higher concentrations than W doping, which is mainly due to the different effect of the dopant in the band structure and position of the Fermi level observed in the Density Functional Theory (DFT) calculations. Interestingly, all Ti-doped films show maximum values of Seebeck coefficient at a range of temperatures of 100-150 °C while W-doped films show this maximum at a higher range of temperatures around 250°C. This behaviour aligns with the described band structures.

The electrical conductivity and power factor evolution with temperature for the $Fe_2VAlTi_x$ and $Fe_2VAlW_x$ alloys for the complete range of doping concentrations is presented in Figure 6. For the Ti-doping case, Figure 6 a), while exhibiting a weak temperature dependence, we can see how the electrical conductivity strongly depends on the amount of Ti, increasing a factor of two by simply adding an Δx=0.1 amount of Ti to the undoped alloy, and with even larger values for x=0.17, then drastically dropping for larger amounts of added Ti. This is possible due to the combined effects of band structure modifications from the gradual substitution of V by Ti in the Heusler lattice, and subsequent changes in carrier concentration and mobility. Ti doping increases the carrier concentration and electrical conductivity. However, the main Fermi surface pocket centred at X point around the Fermi level becomes flatter and more diffuse with increasing Ti content, which reduces hole mobility. Additionally, segregation of V at grain boundaries, especially relevant at higher Ti amounts, may act as electronic scattering centres, reducing the electrical conductivity. Regarding power factor, in Figure 6 b), for the Ti doping case the obtained maximum power factors are in the temperature range of 100-150 °C for all the doping cases. The maximum value obtained is 1300 μW/m·K² for a Ti concentration of x=0.17 at 125 °C.

Switching to W doping, in Figure 6 c) its effect in the electrical conductivity is a 4 to 5-fold increase with respect to the undoped alloy (from 1200-1400 to 5000-5800 S/cm), but is limited to changes of the order of roughly 10% between different W amounts, with a small decrease in values for higher temperatures. On the other hand, going off-stoichiometric largely enhances power factor, Figure 6 d), from maximum values of 600 μW/m·K² for the stoichiometric $Fe_2V_{0.8}W_{0.2}Al$ alloy to more than a 3-fold enhancement (2100 μW/μm·K² at 275°C) for an off-stoichiometry alloy with W concentration of x=0.10. This is due to the mentioned increase in electrical conductivity and Seebeck coefficient.

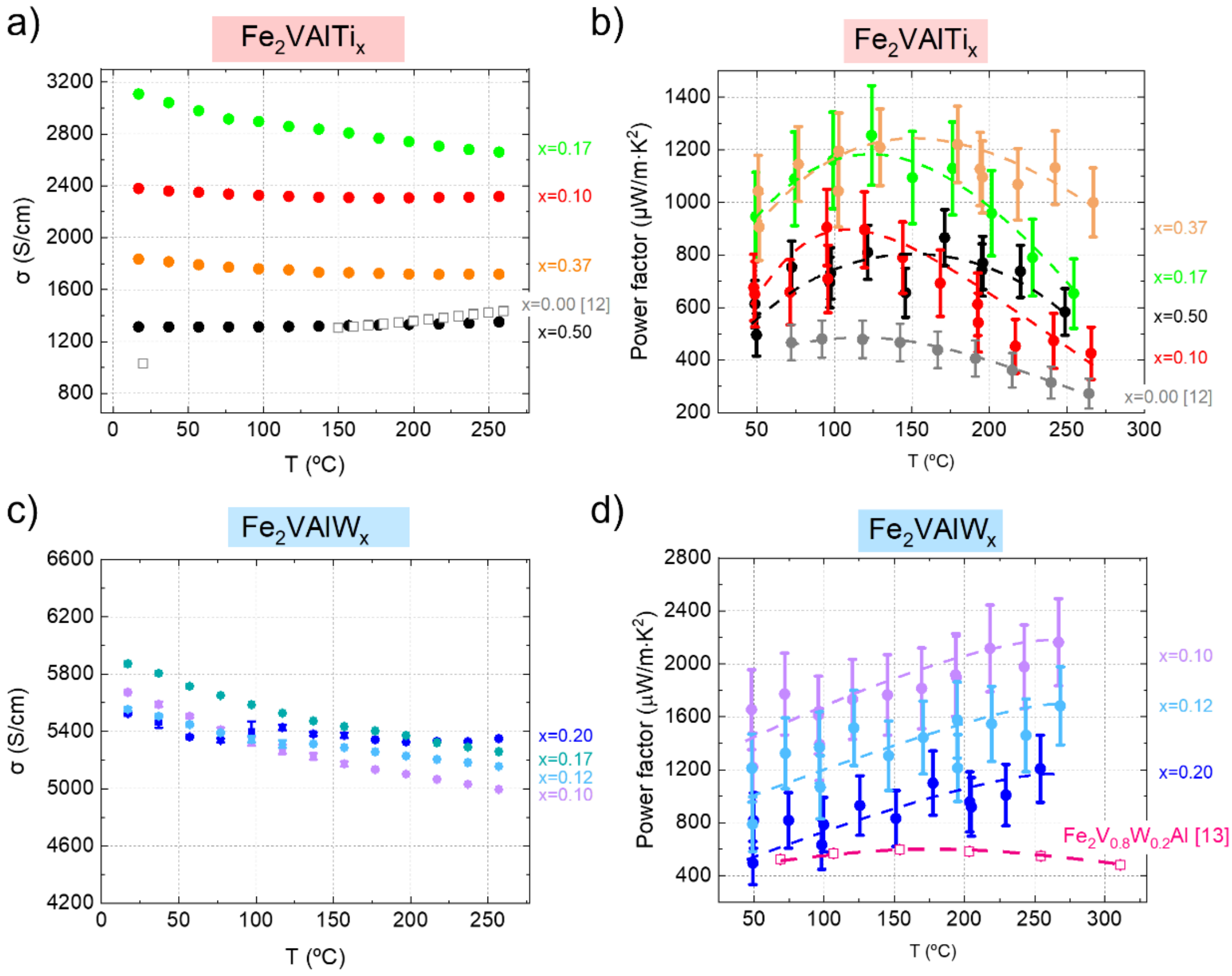

Figure 6: Temperature-dependent electrical conductivity a) and c) and power factor b) and d), respectively of Ti-doped and W-doped films with ranging amounts of Ti and W doping. Error bars of electrical conductivity are smaller than the marker size.

As we have seen, Figures 6 b) and d) highlight the crucial role of fine controlled doping, even in an off-stoichiometric composition, to optimise the material towards maximum PF and, furthermore, it is visible how the use of different doping elements tune the optimum working temperature range. Ti and W doping produce an increment in the temperature of maximum Seebeck coefficient and power factor respect to stoichiometric and $L2_1$-ordered $Fe_2VAl$. The observed detrimental effect of high levels of W doping on Seebeck coefficient translates into the power factor, reducing maximum power factors with increasing amount of W concentration. As it was previously mentioned, this effect is attributed to V and $VW_x$ crystallisation in the alloy.

~~Finally,~~ Figure 7 presents a comparison of thermal transport of films off-stoichiometrically doped with Ti and W with respect to stoichiometric W-doped and undoped films. Figure 7 a) displays the thermal conductivity values ($\kappa$) obtained by TDTR measurements with increasing temperature for undoped, Ti-doped and W-doped $Fe_2VAl$. The different effect of the addition of the two dopants Ti and W is clear. It is evident from the thermal conductivity data that the lighter Ti atom doping has no significant effect on thermal conductivity, whereas heavy W doping significantly reduces the thermal conductivity. Off-stoichiometric W concentration of x=0.12 produces a reduction of 33% at 50 °C. Interestingly, thermal conductivity values of $Fe_2V_{0.8}W_{0.2}Al$ from previous work [13] exhibit lower values, which is explained by the high electrical conductivity in co-deposited films, and subsequent increase of electronic contribution to thermal conductivity, produced by the excess of V and $VW_x$. Figure 7b) shows the effect of the dopants in the lattice thermal transport. It is visible that W doping, for both off-stoichiometric and stoichiometric cases, reduces notably the lattice thermal conductivity (more than two-fold decrease at 50°C) while Ti-doped films exhibit the same value as undoped $Fe_2VAl$. Despite exhibiting different electronic conductivities, it is observed that stoichiometric $Fe_2V_{0.8}W_{0.2}Al$ and off-stoichiometric $Fe_2VAlW_{0.12}$ exhibit similar values of lattice thermal conductivity. It is noted that all samples exhibit nearly constant behaviour of their lattice thermal conductivities with temperature, this behaviour aligns with previously reported data of $Fe_2VAl$ [42,43].

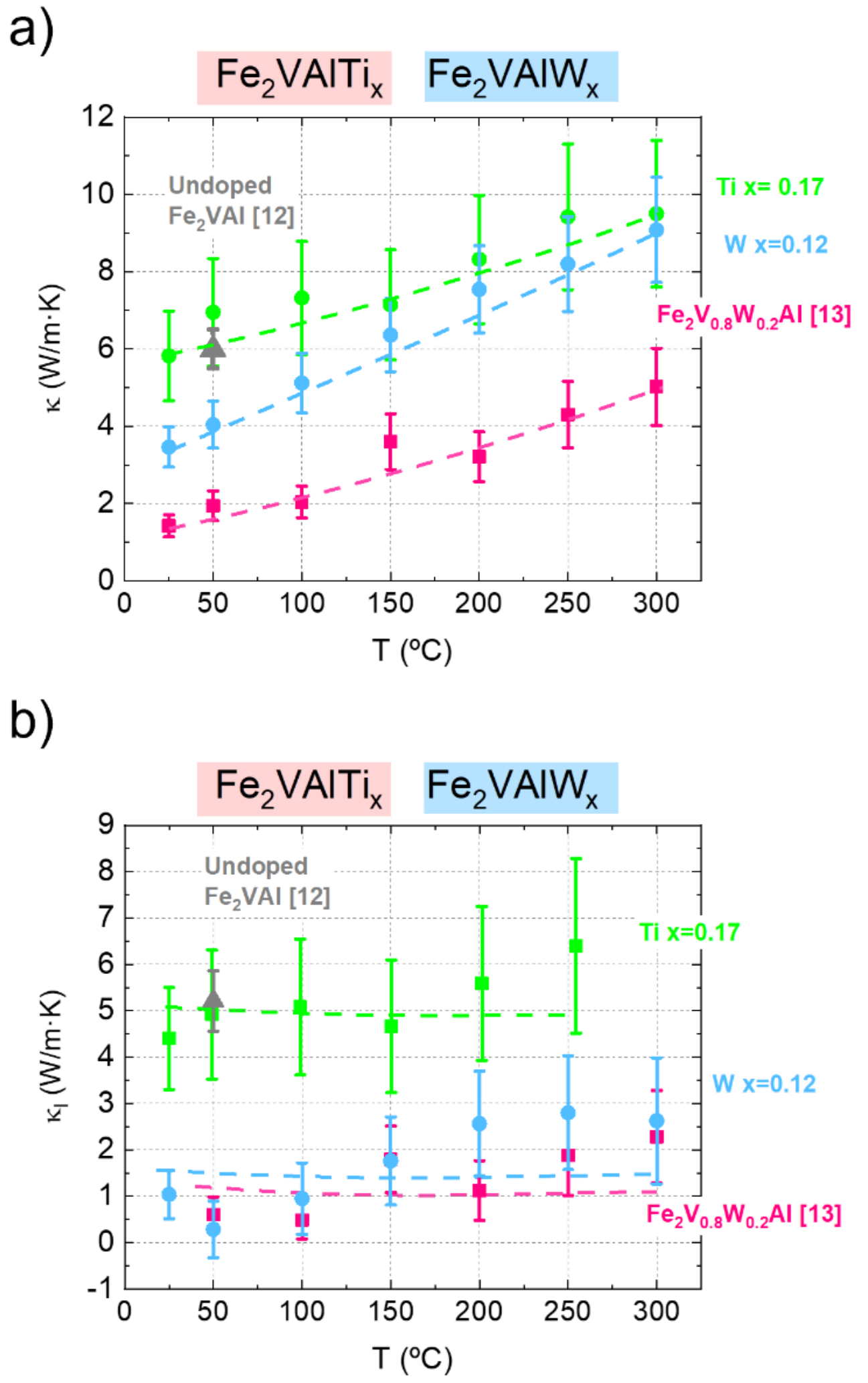

Figure 7: Temperature-dependent thermal conductivity a) and lattice thermal conductivity b) for representative films with Ti and W doping. Total and lattice thermal conductivity of undoped $Fe_2VAl$ and $Fe_2V_{0.8}W_{0.2}Al$ are plotted for comparison reasons and were obtained from [12,13].

It should be noted that for W-doped films, which exhibit confirmed two-carrier (electron + hole) transport, a bipolar thermal conductivity contribution $\kappa_{bip} \propto [(\sigma_e \cdot \sigma_h)/(\sigma_e+\sigma_h)^2]\,(S_e-S_h)^2T$ is expected in addition to the Wiedemann–Franz electronic term. Based on the measured conductivities and Seebeck values, this contribution is estimated to be below 7% of the total $\kappa$ at the temperatures of maximum zT, and is therefore treated as a minor correction here.

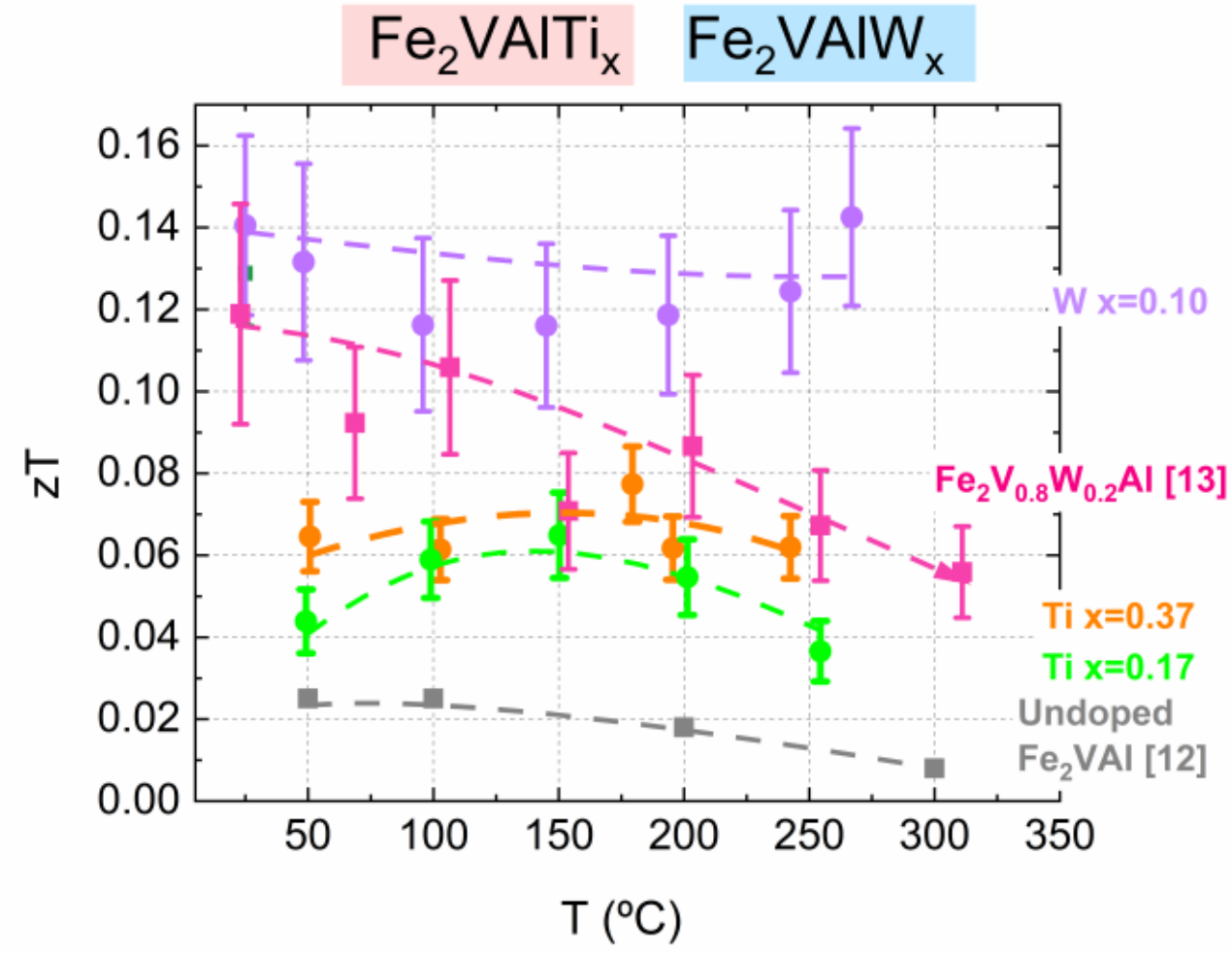

Figure 8: Temperature-dependent zT for representative films with Ti and W doping. zT of undoped $Fe_2VAl$ and $Fe_2V_{0.8}W_{0.2}Al$ were obtained from [12,13].

The differing thermal transport properties of W-doped and Ti-doped films lead to significant variations in *zT*. Figure 8 shows the temperature-dependent *zT* for undoped, Ti-doped and W-doped thin films. Ti-doped films present a more than three-fold enhancement of maximum *zT* with respect to undoped $Fe_2VAl$, achieving maximum *zT* of 0.08 at 175 °C on a film with a Ti concentration of x=0.37, mainly due to the power factor enhancement produced by Ti doping. On the other hand, W-doping results in an even larger enhancement in *zT*, as W is an element with high atomic number that produces large phonon scattering. The stoichiometrically W-doped $Fe_2V_{0.8}W_{0.2}Al$ film from our previous work[13] presents a maximum *zT* of 0.12 at room temperature, accounting for a more than fourfold increase with respect the undoped $Fe_2VAl$. The off-stoichiometrically doped film presented in this work further enhances the maximum *zT* with respect to the $Fe_2V_{0.8}W_{0.2}Al$ film obtaining a maximum of *zT*=0.14±0.02 at RT and 270 °C on a film with W concentration of x=0.1. It is observed that off-stoichiometric doping produces a larger power factor and an also higher thermal conductivity compared with stoichiometric $Fe_2V_{0.8}W_{0.2}Al$ from [13] that overall results in a slightly larger figure of merit value that is closer to maximum *zT* values obtained in W-doped $Fe_2VAl$ bulk alloys [14,15].

## Conclusions

The combined structural, transport and thermoelectric results of this work describe simple design guidelines for $Fe_2VAl$-based thermoelectric thin films p or n type with outstanding power factors. $L2_1$ chemical order is a prerequisite for achieving large Seebeck coefficients and enhanced power factors in both undoped and doped films, thanks to the electronic structure modifications induced by this order. Combined theoretical calculations and experimental characterisation reveal different doping mechanisms in Ti and W doping. Off-stoichiometrically added Ti atoms replace effectively V in the alloy, tuning the Fermi level into the valence band and reaching optimal p-type thermoelectric performance at intermediate Ti contents, where the Seebeck coefficient and electrical conductivity are maximised. Off-stoichiometrically added W atoms, in turn, compete with V atoms to occupy in the lattice sites, their leftovers resulting in V and $VW_x$ aggregates crystallisation. W off-stoichiometric doping produces a band shift towards lower energy sates providing n-type behaviour and a band crossing at the $\Gamma$ point that results in asymmetric multi-carrier transport. In terms of thermoelectric performance, W doping enhances the Seebeck coefficient from stoichiometric $Fe_2V_{0.8}W_{0.2}Al$ and undoped $Fe_2VAl$ values, and more notably the electrical conductivity, resulting in a more than two-fold enhancement in the power factor maximum value. The larger atomic number of W attains for a greater reduction of thermal conductivity comparing with Ti, achieving W off-stoichiometrically doped films the greatest zT value reported in this work of 0.14. Overall, co-deposition-assisted off-stoichiometric doping enhances the power factor and zT, p and n-type, of $Fe_2VAl$, in a single-step process compatible with industrial magnetron sputtering, though the 900°C deposition temperature currently limits substrate choice to thermally stable oxide ceramics, and future work targeting reduced deposition temperatures through, e.g., rapid thermal annealing protocols could extend applicability to broader substrate platforms.

More broadly, high-temperature co-deposition combined with controlled off-stoichiometry appears as a versatile strategy to engineer the balance between band-edge electronic structure and scattering mechanisms in full Heusler thermoelectrics based on earth-abundant, non-toxic elements.

## Conflicts of interest

There are no conflicts to declare.

## Data availability

All the data supporting the findings will be available at https://digital.csic.es/ after publication. Additional data are available from the corresponding author upon reasonable request.

## Acknowledgements

The authors would also like to acknowledge the service from the MiNa Laboratory at IMN, and its funding from CM (project SpaceTec, S2013/ICE2822), MINECO (project CSIC13-4E-1794), and EU (FEDER, FSE). This work was funded by projects THERMHEUS grant TED2021-131746B-I00 funded by MICIU/AEI/10.13039/501100011033 and by the "European Union NextGenerationEU/PRTR" and ERC Adv. POWERbyU grant agreements ID: 101052603 Founded by European Research Council (ERC), grant PID2022-138063OB-I00 funded by MICIU/AEI/10.13039/501100011033 and by FEDER, UE. We thankfully acknowledge the computer resources at Lusitania (Cenits-COMPUTAEX), Red Española de Supercomputación, RES (QHS-2023-1-0028) and Albaicín (Centro de Servicios de Informática y Redes de Comunicaciones - CSIRC, Universidad de Granada). KL acknowledges Aid JDC2023-050703-I funded by MICIU/AEI/10.13039/501100011033 and by the ESF+.

*SUPPLEMENTARY INFORMATION*

# Off-stoichiometric variable doping for exceptional power factors in L2$_1$ $Fe_2VAlM_x$ (M=Ti, W) epitaxial thin films.

Jose María Domínguez-Vázquez[a], Miguel Angel Tenaguillo[a], Ketan Lohani[a], Olga Caballero-Calero[a], Jose J. Plata[b], Antonio M. Marquez[b], Alfonso Cebollada[a], Andrés Conca[a,*], Ernst Bauer[c] and Marisol Martín-González[a]

[a]Instituto de Micro y Nanotecnología, IMN-CNM, CSIC (CEI UAM+CSIC), Isaac Newton 8, E-28760 Tres Cantos, Madrid, Spain
[b]Dpto de Química Física, Facultad de Química, Universidad de Sevilla, Sevilla (Spain).
[c]Institute of Solid-State Physics, Technische Universität WienWiedner, Hauptstraße 8-10, Vienna, 1040, Austria
*Corresponding author: andres.conca@csic.es

Figure SI 1 shows the off-specular characterisation of Fe2VAlMx (M=Ti, W) Figure SI 2 displays the calibration plots of Ti and W concentration with electrical power applied to the magnetron. Table 1 shows the formation energy of all the possible compounds taking into account interstitial, substitution and interstitial + substitution dynamics. Table 2 displays the calculated lattice parameters for different Ti and W concentrations. Figure SI 3 shows the Energy-dispersive X-ray Spectroscopy (EDX) maps for the two films with highest amount of Ti and W doping. Figure SI 4 shows the measured temperature-dependent mobilities of Ti-doped $Fe_2VAl$ thin films for different Ti concentrations. Figure SI 5 depicts the temperature-dependent electron a) and hole b) mobilities of $Fe_2VAlW_x$ for concentrations x= 0.12 and 0.20.

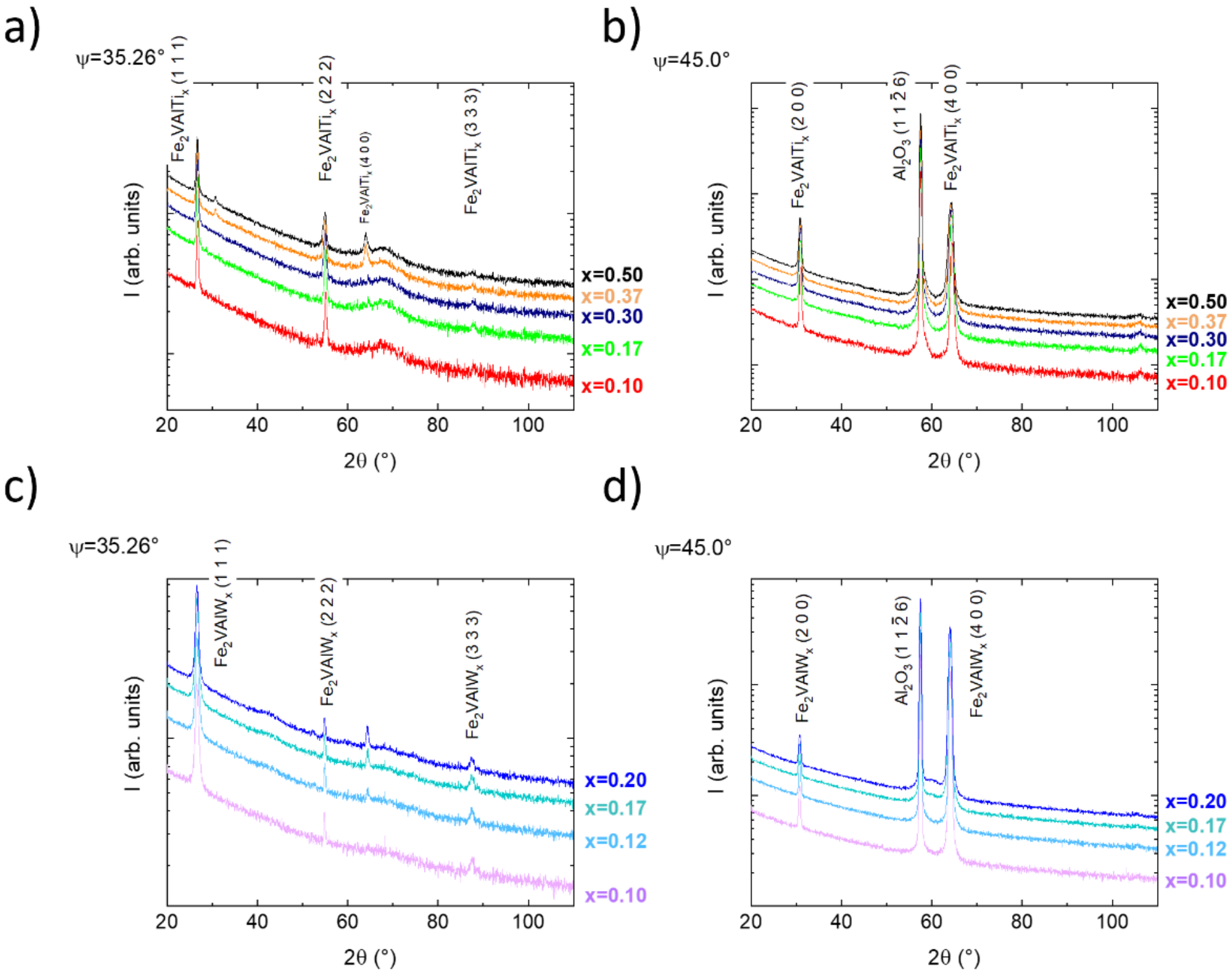

*SI 1:Off-specular X-ray diffraction (XRD) θ-2θ measurements of $Fe_2VAlM_X$ (M=Ti, W) showing the high order diffraction peaks of (1 1 1) and (2 0 0) for the complete set of films.*

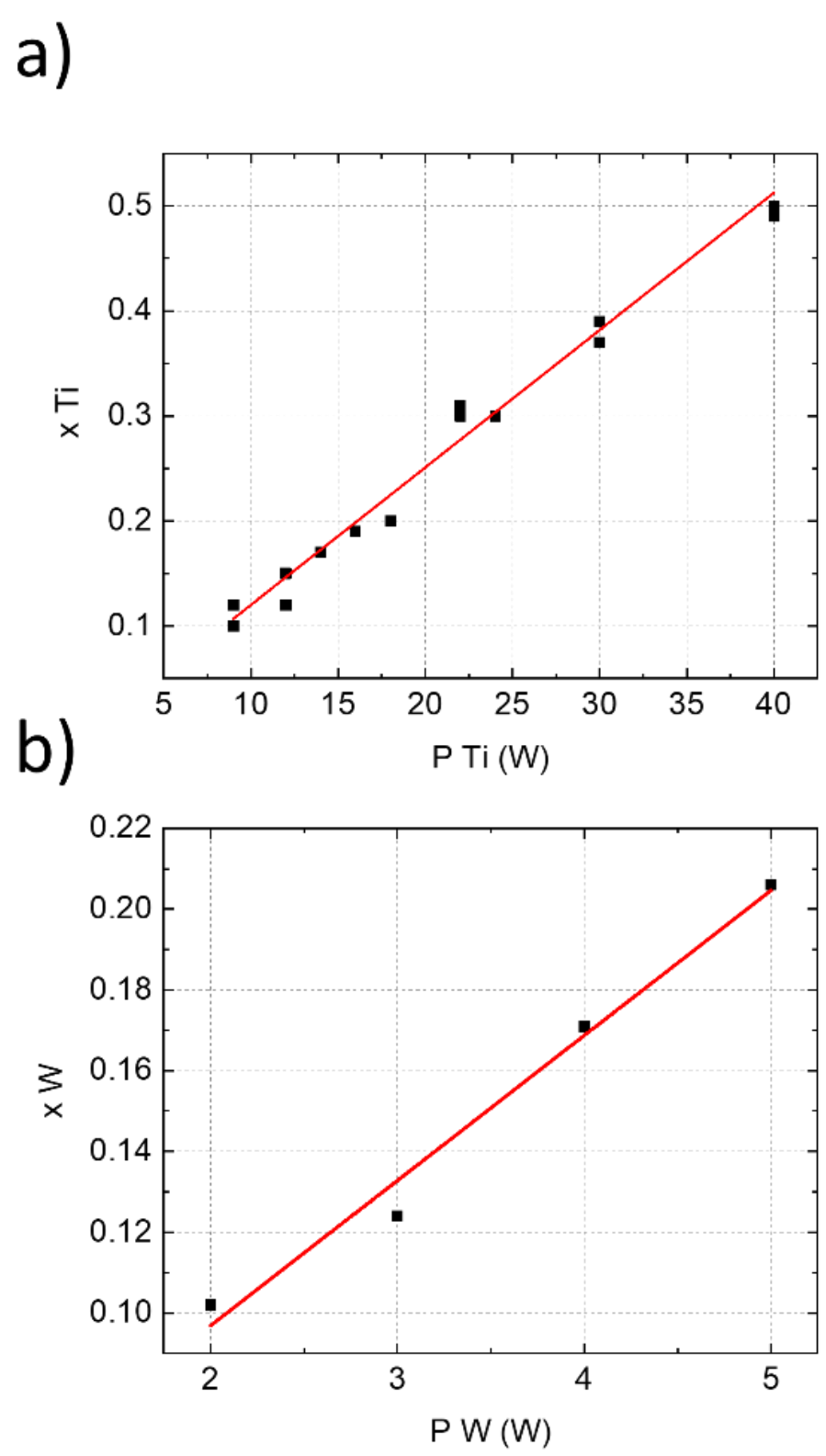

*SI 2: Calibration of a) Ti and b) W concentration with electrical power applied to the magnetron.*

| W | | | | | | |
|---|---|---|---|---|---|---|
| | **Interstitial** | **Substitution** | **Interstitial + substitution** | | | |
| | | | **1** | **2** | **3** | **4** |
| **Al** | 6.65 | 2.29 | 7.16 | 8.13 | 7.06 | 6.54 |
| **V** | | **0.90** | 7.33 | 6.55 | 6.90 | 7.38 |
| **Fe** | | 2.10 | 7.07 | 5.69 | 5.39 | 5.96 |
| Ti | | | | | | |
| | **Interstitial** | **Substitution** | **Interstitial + substitution** | | | |
| | | | **1** | **2** | **3** | **4** |
| **Al** | 6.06 | 0.80 | 4.74 | 5.64 | 6.09 | 6.30 |
| **V** | | **-0.15** | 6.57 | 6.32 | 6.01 | 5.87 |
| **Fe** | | 1.71 | 5.97 | 5.17 | 4.89 | 5.30 |

*Table 1 : Calculated formation energies (eV) of $Fe_2VAlM_x$ compounds with M=Ti,W. Interstitial, substitution and interstitial + substitution dynamics are explored. Interstitial +substitution scenario is specified for the 4 possible interstitial positions of the expelled atom (Al, V or Fe).*

| Ti% | a=b=c (Å) | W% | a=b=c (Å) |
|---|---|---|---|
| 9 | 5.717 | 9 | 5.719 |
| 19 | 5.728 | 13 | 5.724 |
| 28 | 5.738 | 16 | 5.729 |
| 38 | 5.752 | 19 | 5.734 |
| 47 | 5.758 | | |

*Table 2: Calculated lattice parameters for different Ti and W concentrations.*

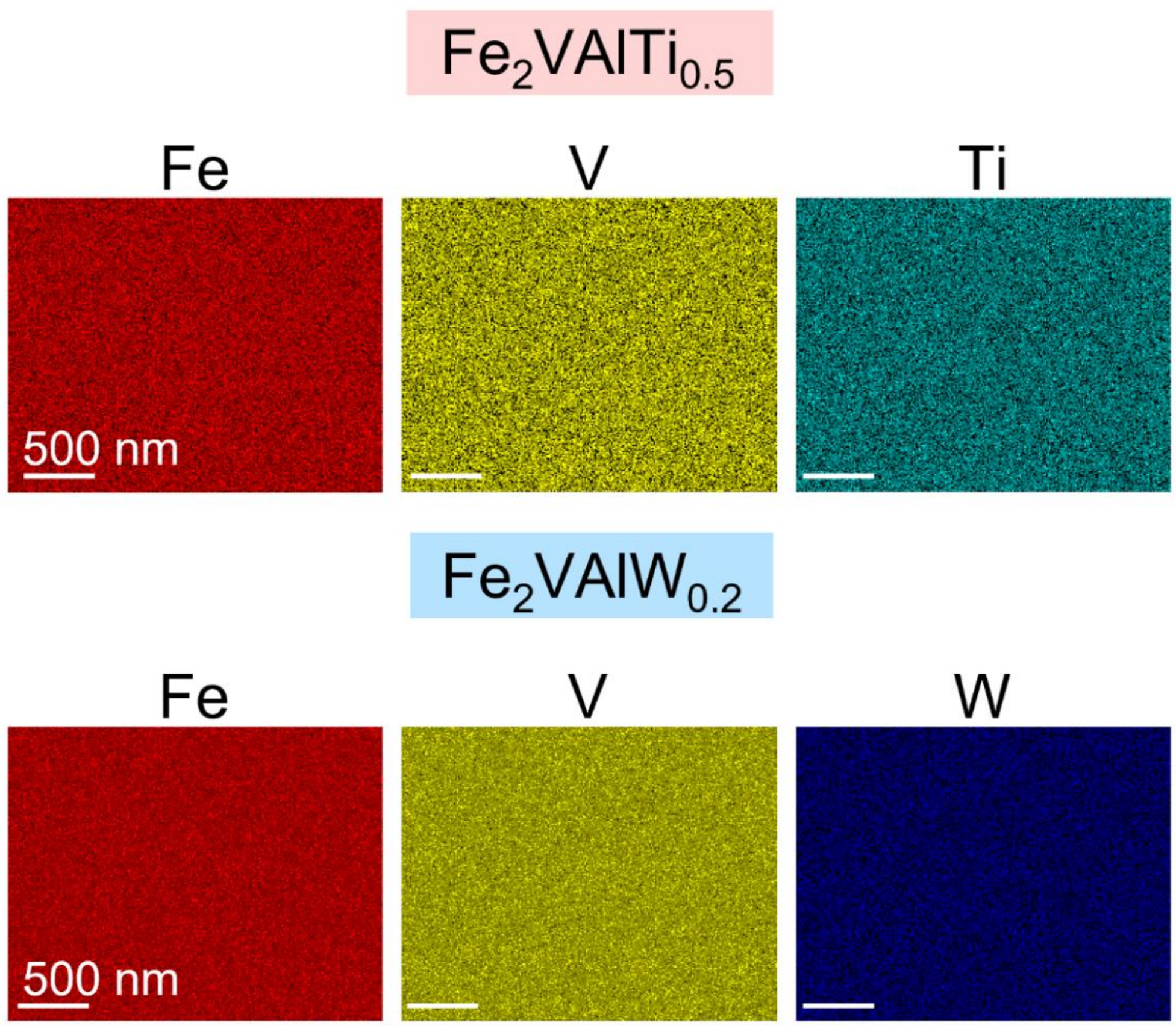

*SI 3: Energy-Dispersive X-ray Spectroscopy of highest Ti-doped and W-doped Fe2VAl thin films. Al signal is not portrayed as it is present also in the $Al_2O_3$ substrate.*

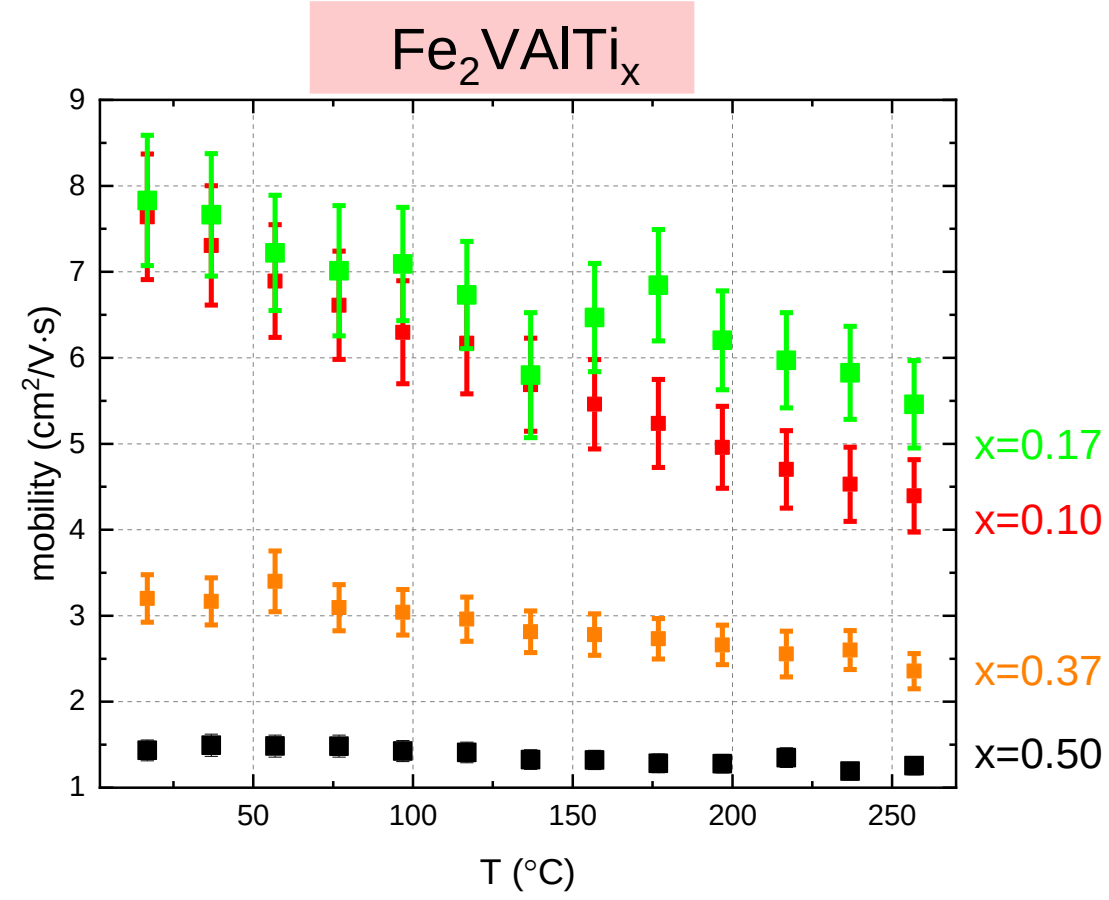

*SI 4: Temperature-dependent carrier mobility of $Fe_2VAlTi_x$ thin films with different Ti concentrations.*

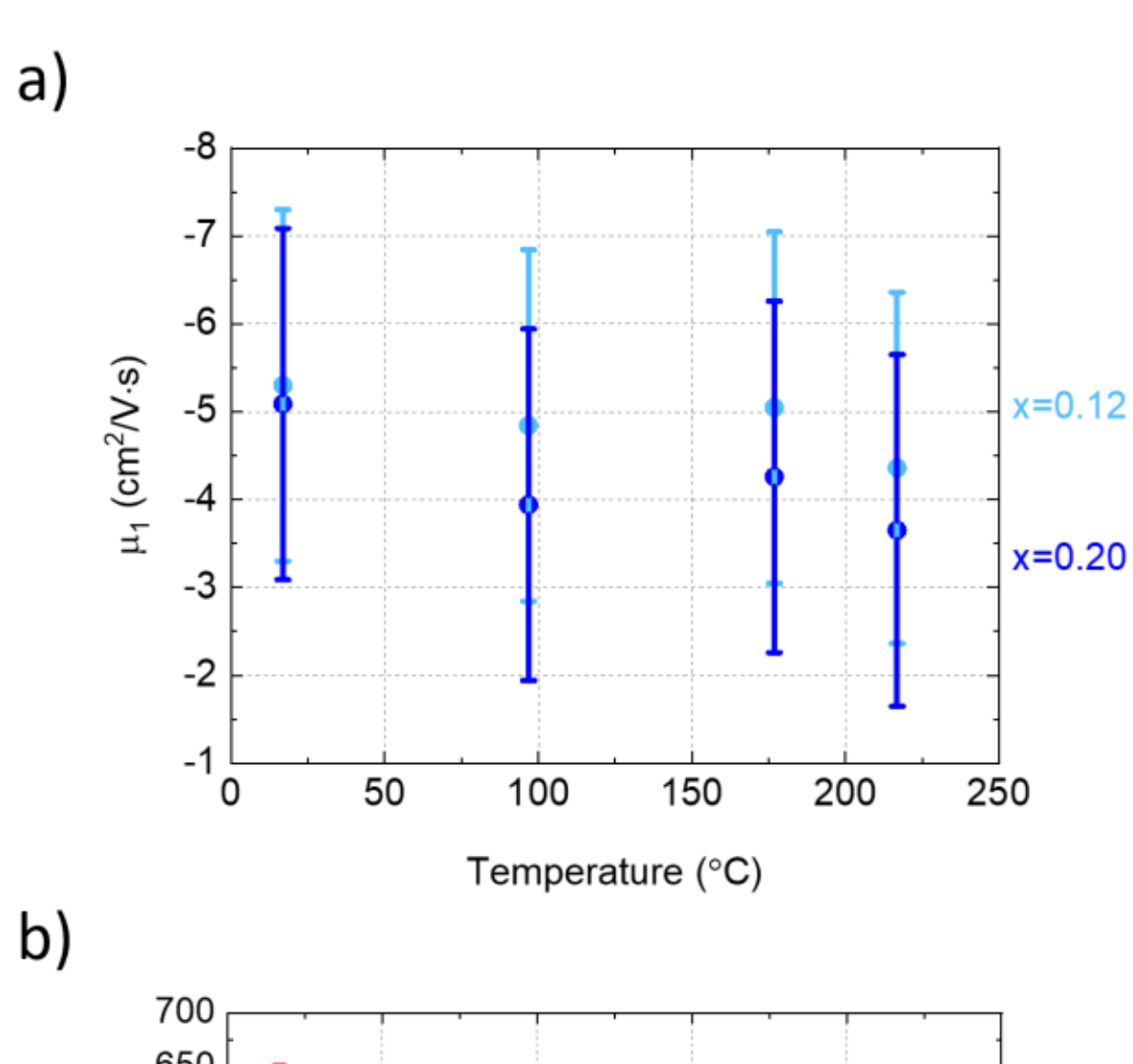

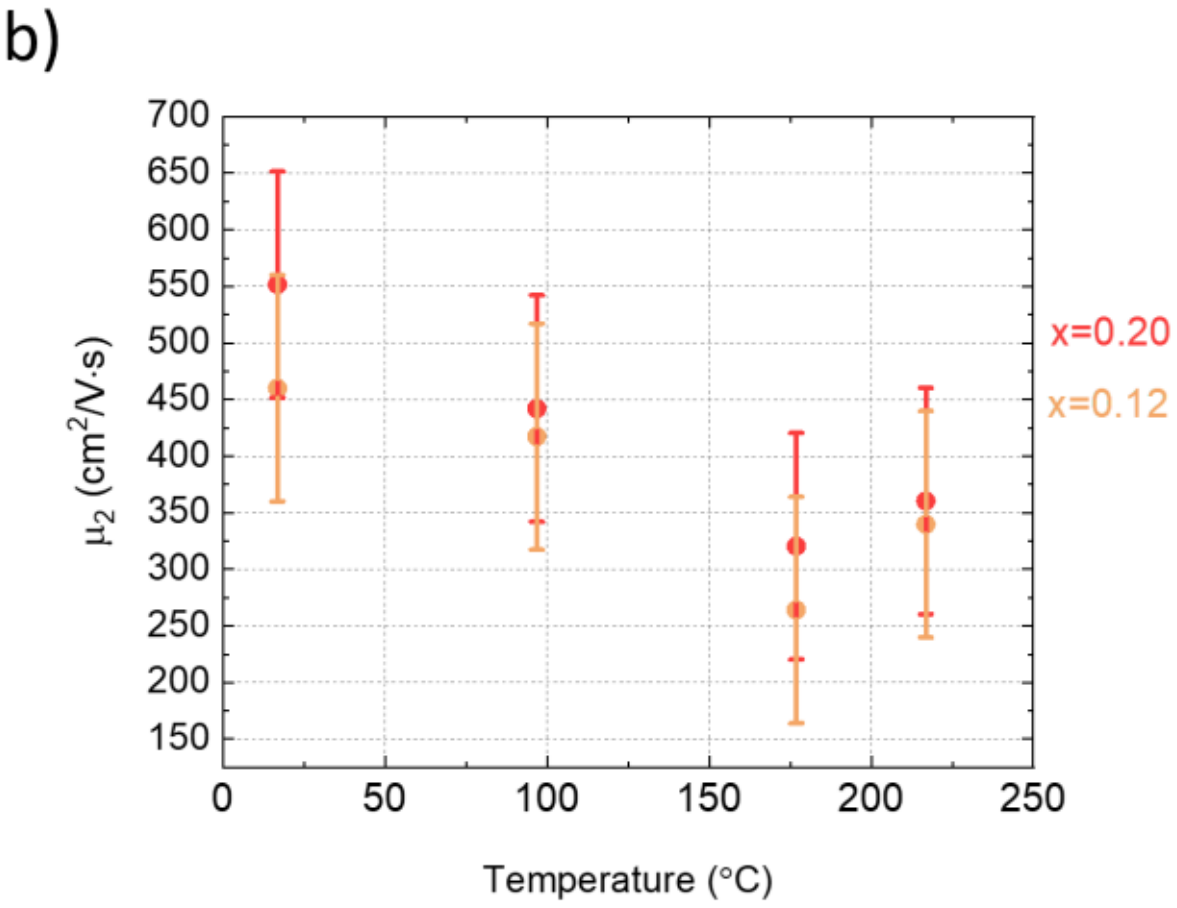

*SI 5: Temperature-dependent electron a) and hole b) mobilities of $Fe_2VAlW_x$ for concentrations x=0.12 and 0.20.*

## Reproducibility statements

This work consists of two deposited series (simultaneously depositing on $Al_2O_3$ and MgO substrates) with Ti and W doping, and a fixed deposition temperature ($T_{dep}$) of 900°C. In order to avoid temporal drifts effects in the deposition system each sample series was deposited in a random order of Ti or W concentration (i.e., chronologically the samples were deposited with a doping concentration covering in a random fashion the doping concentration range). The whole deposition spanned for two and a half months, showing the stability of the process. In order to confirm previous demonstrations of $L2_1$ ordering enhancement of thermoelectric properties, a Ti-doped sample was deposited with x=0.17 at a deposition temperature of $T_{dep}$=550°C crystallising the B2 structure, this film showed a Seebeck coefficient of +7 μV/K. These considerations ensure the reproducibility and stability of the deposition procedure.